\documentclass[aps,twocolumn]{revtex4}
\usepackage{bm}
\usepackage{amsmath}
\usepackage{graphicx}
\usepackage{subfigure}
\usepackage[usenames,dvipsnames]{color}
\definecolor{darkblue}{RGB}{0,0,196}
\usepackage[colorlinks=true,linkcolor=darkblue,citecolor=darkblue,urlcolor=darkblue]{hyperref}
\usepackage{setspace}
\usepackage{hyperref}
\usepackage{xcolor}
\hypersetup{
  colorlinks   = true, 
  urlcolor     = red, 
  linkcolor    = blue, 
  citecolor   = blue 
}
\usepackage{footmisc}
\usepackage[makeroom]{cancel}
\usepackage{comment}
\def\be{\begin{equation}}
\def\ee{\end{equation}}
\def\ba{\begin{eqnarray}}
\def\ea{\end{eqnarray}}
\usepackage{graphicx}
\usepackage{amsmath,bbm}
\usepackage{amssymb,bm}
\usepackage{yfonts}

\begin{document}
\title{Estimating Elliptic Flow Coefficient in Heavy Ion Collisions using Deep Learning}
\author{Neelkamal Mallick$^{1}$}
\author{Suraj Prasad$^{1}$}
\author{Aditya Nath Mishra$^2$}
\author{Raghunath Sahoo$^{1,3,}$\footnote{Corresponding author: $Raghunath.Sahoo@cern.ch$}}
\author{Gergely G\'abor Barnaf\"oldi$^2$}
\affiliation{$^{1}$Department of Physics, Indian Institute of Technology Indore, Simrol, Indore 453552, India}
\affiliation{$^{2}$Wigner Research Center for Physics, 29-33 Konkoly-Thege Mikl\'os Str., H-1121 Budapest, Hungary}
\affiliation{$^{3}$CERN, CH 1211, Geneva 23, Switzerland}

\begin{abstract}
Machine Learning (ML) techniques have been employed for the high energy physics (HEP) community since the early 80s to deal with a broad spectrum of problems. This work explores the prospects of using Deep Learning techniques to estimate elliptic flow ($v_2$) in heavy-ion collisions at the RHIC and LHC energies. A novel method is developed to process the input observables from particle kinematic information. The proposed DNN model is trained with Pb-Pb collisions at $\sqrt{s_{\rm NN}} = 5.02$ TeV minimum bias events simulated with AMPT model. The predictions from the ML technique are compared to both simulation and experiment. The Deep Learning model seems to preserve the centrality and energy dependence of $v_2$ for the LHC and RHIC energies. The DNN model is also quite successful in predicting the $p_{\rm T}$ dependence of $v_2$. When subjected to event simulation with additional noise, the proposed DNN model still keeps the robustness and prediction accuracy intact up to a reasonable extent.

\end{abstract}
\date{\today}
\maketitle 

\section{Introduction}
\label{intro}
Extremely hot and dense state of the strongly interacting matter is being investigated in collisions of heavy nuclei at ultra-relativistic energies for decades. In high-energy particle colliders, it is believed, that the state of the early Universe can be recreated in tiny volumes of the order of a few cubic fermi (${\rm fm}^3$). This state of the deconfined color partons is called as Quark-Gluon Plasma (QGP). Due to the very short duration of its existence, direct observation of QGP is not possible, however, signatures of the evidence can be measured. 

The presence of transverse collective expansion ({\it i.e.} transverse flow) in strongly coupled dense nuclear matter formed in relativistic heavy-ion collisions serves as an important signature for QGP~\cite{Heinz:2013th}. In particular, the transverse flow is anisotropic in nature meaning particle emission is anisotropic in the momentum space. Anisotropic flow of different order could be expressed as Fourier expansion
coefficients of produced particles' azimuthal distribution in
the momentum space. Experiments observe the existence of finite anisotropic flow, mainly the elliptic flow ($v_2$) in heavy-ion collisions~\cite{STAR:2003wqp,ALICE:2010suc,ALICE:2011ab,ALICE:2014dwt}. For decades, investigations are being performed to solve the equations of relativistic fluid dynamics to theoretically model the elliptic flow. Although, these theoretical models based on relativistic fluid dynamics were successful in presenting some of the low-{$p_{\rm T}$} phenomena, it over-predicted the observed elliptic flow~\cite{Csanad:2005gv,Kolb:2000fha}. The solution appeared in the form of hybrid models that couple both (ideal) fluid dynamics applied to the QGP phase and kinetic descriptions applied to microscopic hadron cascade phase~\cite{Bass:2000ib,Nonaka:2006yn,
Teaney:2001av,Hirano:2005xf}. In intermediate to high-$p_{\rm T}$, elliptic flow suffers combined consequences from jet quenching and parton energy loss due to the dense medium and thus, simple hydro models fail to explain the data~\cite{Kolb:2000fha,PHOBOS:2004vcu}.  A recent study has presented the implementation of hydro freeze-out at low-$p_{\rm T}$, parton coalescence at intermediate-$p_{\rm T}$, and fragmentation at high-$p_{\rm T}$ along with coupled linear Boltzmann transport-hydro model \cite{Zhao:2021vmu}.  This could explain both $R_{\rm AA}$ and $v_2$  simultaneously from low to intermediate and high-$p_{\rm T}$ in high-energy heavy-ion collisions. 

Apart from understanding various theoretical aspects of elliptic flow, there are challenges in estimating $v_2$ in experiments. By definition, $v_2$ requires the information of the reaction plane angle on an event-by-event basis, whose measurement is non-trivial in experiments. There are couple of methods that offer the solution such as the complex reaction plane identification
method~\cite{Poskanzer:1998yz}, the cumulant method~\cite{Borghini:2000sa}, and the Lee-Yang zeroes method~\cite{Bhalerao:2003xf}. There are even attempts to implement the Principal Component Analysis (PCA) methods to estimate $v_2$ as well~\cite{Bhalerao:2014mua,Hippert:2019swu,CMS:2017mzx,
Gardim:2019iah}. For the first time, we propose a Deep Learning method in the Machine Learning framework to estimate $v_2$.
Machine Learning (ML) techniques have been employed for the high energy physics (HEP) community since the early 80s to deal with a broad spectrum of problems~\cite{Carleo:2019ptp, Albertsson:2018maf,
Ortiz:2020rwg, Radovic:2018dip}. With the advancement of superior hardware and smart algorithms, ML has become the most popular tool for statistical, data-driven, and prediction-based applications. It has the ability to perceive unique features and patterns in data to solve unconventional problems such as classification, regression, and clustering, just to name a few.

The motivation of this study is to explore the prospects of using Deep Learning techniques to estimate elliptic flow. We also proceed to show that the model is capable of learning centrality and transverse momentum dependence of $v_2$ as well. A simple attempt is made in this paper to estimate elliptic flow in heavy-ion collisions event-by-event from various particle kinematic information using a feed-forward Deep Neural Network (DNN). The proposed DNN model is trained with minimum bias Pb-Pb collisions at $\sqrt{s_{\rm NN}} = 5.02$ TeV simulated with AMPT model. The predictions of the machine are compared to both simulation and experiment. Although, the training of the model is a bit CPU expensive, the trained model could be applied independently to estimate $v_2$, which is faster and economical.  

This paper is organised as follows. We begin with a brief introduction to event generation using AMPT model and the target observable in Sec.~\ref{Data}. In Sec.~\ref{ML}, we describe the proposed ML regression model based on Deep Learning in detail and provide some quality assurance plots along with estimation of systematic uncertainties. Finally, the results are presented in Sec.~\ref{results} and we summarise our findings in Sec.~\ref{summary}. 

\begin{figure*}[ht!]
   \includegraphics[scale=0.25]{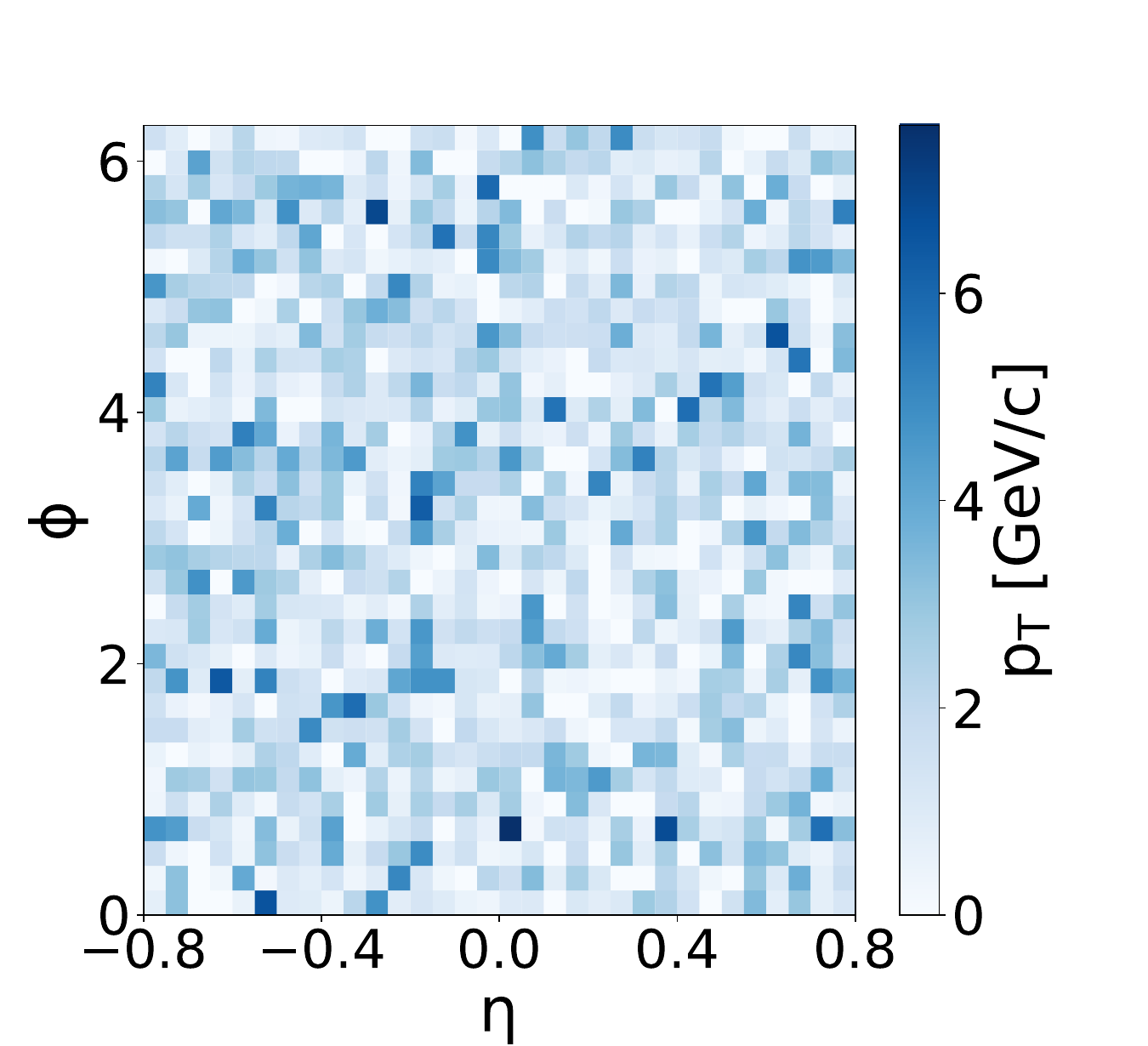}
   \includegraphics[scale=0.25]{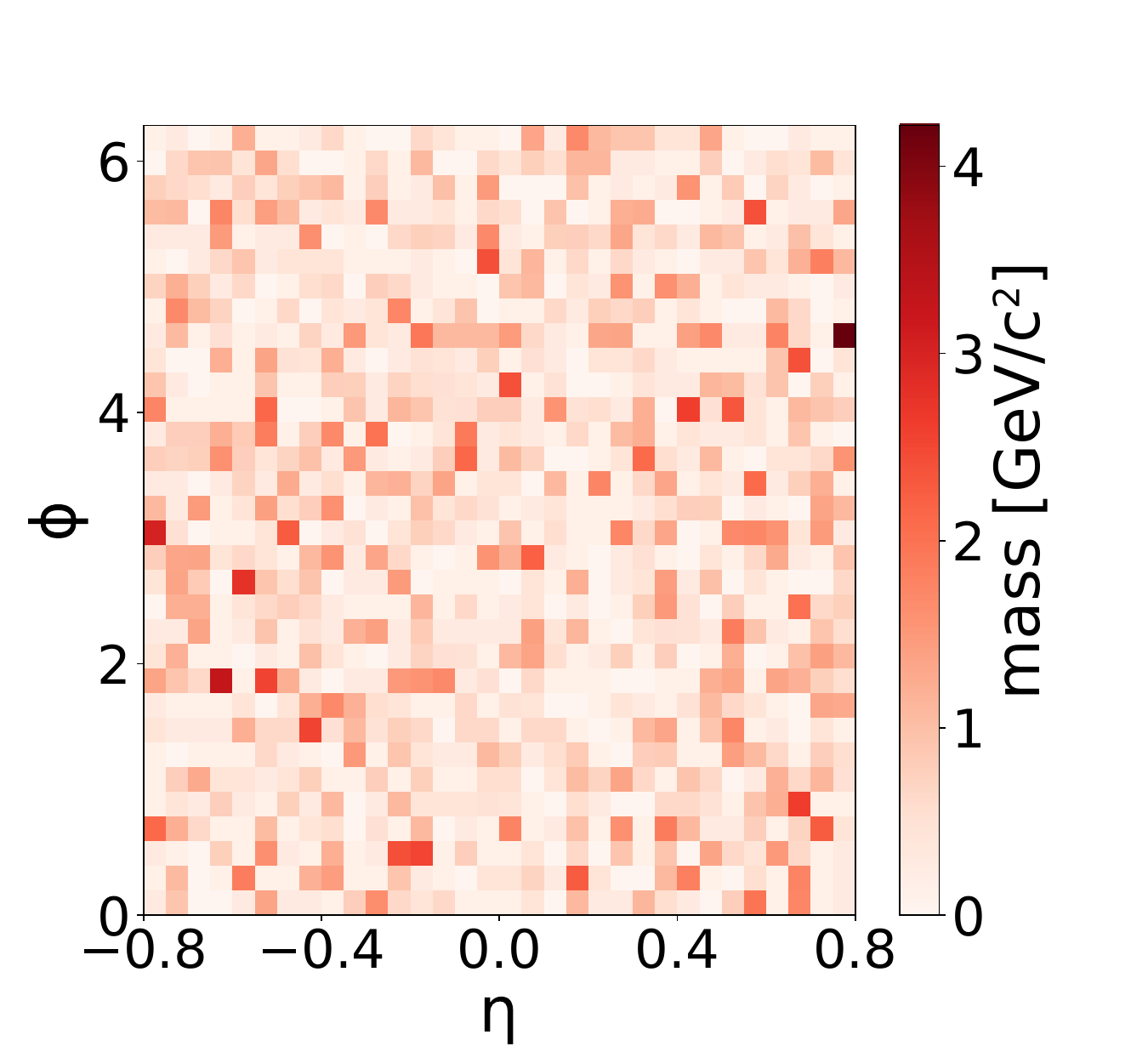}
   \includegraphics[scale=0.25]{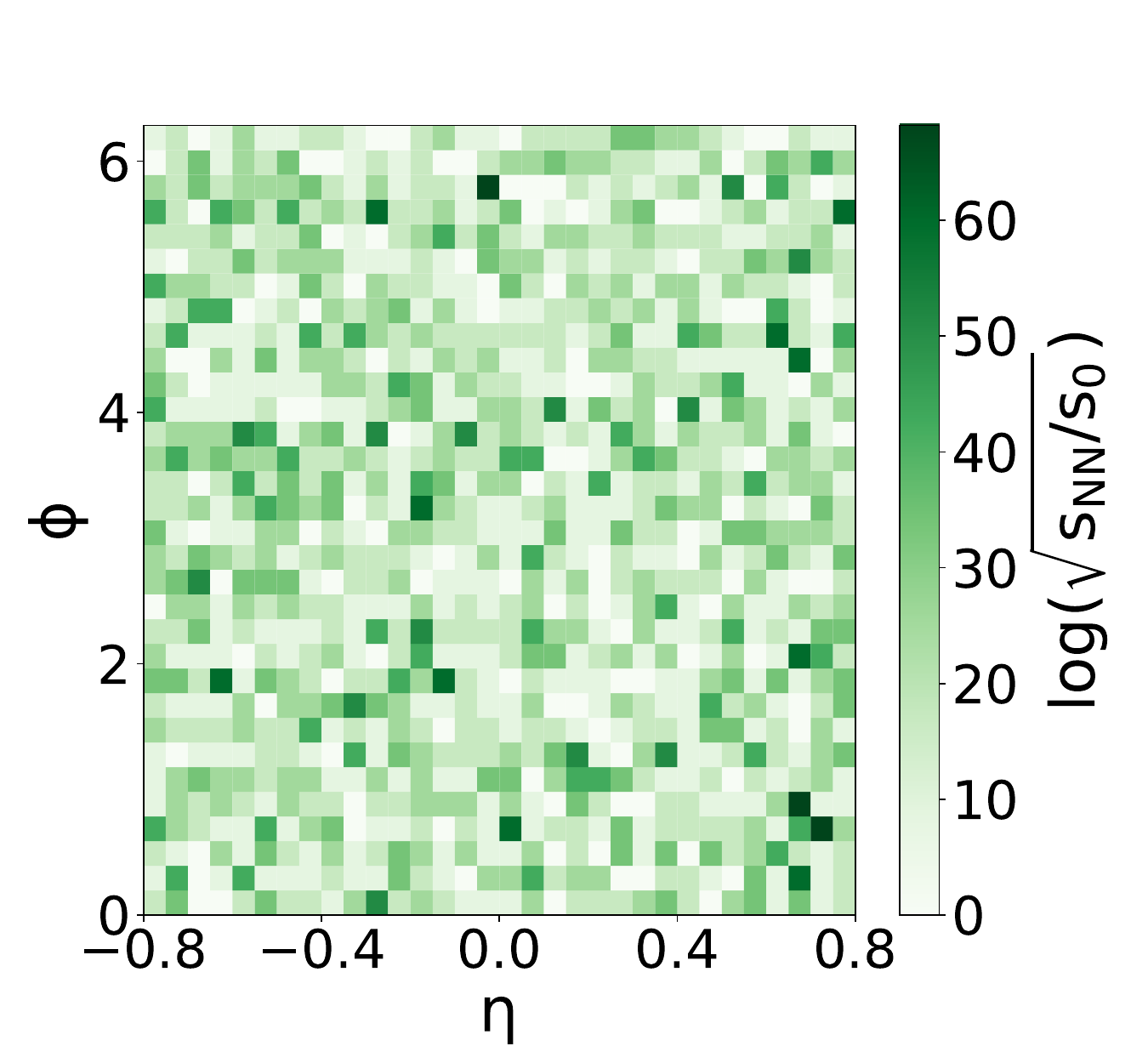}
    \caption{(Color online) The $(\eta {\rm -}\phi)$~space with $32\times32$ bins showing the three layers of information for a single minimum bias Pb-Pb collision at $\sqrt{s_{\rm NN}} = 5.02$~TeV from AMPT model. The $p_{\rm T}$, mass and $\log(\sqrt{s_{\rm NN}/s_0})$ weighted plots are shown in blue, red and green colours respectively.}
    \label{fig1}
\end{figure*}

\section{Event Generation and target observable}
\label{Data}
In this section, we describe briefly the event generation using a multi-phase transport model and then define the target observable the elliptic flow.
\subsection{AMPT Model}
\label{AMPTModel}
A multi-phase transport model (AMPT) is a Monte Carlo event generator for simulating p-A and A-A collisions from RHIC to LHC energies to study the properties of hot and dense nuclear matter~\cite{Lin:2004en}. It has four main components, namely, the fluctuating initial conditions, followed by the parton cascade, hadronization mechanism, and hadron cascade. These are discussed below.
\begin{itemize}
    \item Initialization of collision is done by obtaining the spatial and momentum distributions of the hard minijet partons and soft string excitations from the HIJING model~\cite{ampthijing}. The inbuilt Glauber model is used to calculate and convert the cross-section of the produced mini-jets in pp collisions to heavy-ion collisions.
    \item Zhang's parton cascade (ZPC) model is used to perform the partonic interactions and parton cascade which currently includes the two-body scatterings with cross-sections obtained from the pQCD with screening masses~\cite{amptzpc}.
    \item Hadronization mechanism includes the default mode where the Lund string fragmentation model is used to recombine the partons with their parent strings and then the strings are converted to hadrons, whereas, in the string melting mode the transported partons are hadronized using a quark coalescence mechanism~\cite{Lin:2001zk,He:2017tla}.
    \item Finally, the scattering among the produced hadrons are performed using a relativistic transport model (ART) by meson-meson, meson-baryon and baryon-baryon interactions~\cite{amptart1,amptart2}.
    
\end{itemize}

We have used the AMPT string melting mode (AMPT version 2.26t9b) for event generation as the anisotropic flow and particle spectra in intermediate-$p_{\rm T}$ region is well explained by the quark coalescence mechanism for hadronization~\cite{ampthadron2,ampthadron3,Greco:2003mm}. The AMPT settings in the current work are the same as those reported in Refs.~\cite{Tripathy:2018bib,Mallick:2020ium, Mallick:2021wop} for heavy-ion collisions at the LHC energies. For the definition of centrality of collisions, we have followed Ref.~\cite{Loizides:2017ack}. We have simulated minimum bias 200K events for Pb-Pb collisions at $\sqrt{s_{\rm NN}} = 5.02$ TeV, 100K events for Pb-Pb collisions at $\sqrt{s_{\rm NN}} = 2.76$ TeV, and 100K events for Au-Au collisions at $\sqrt{s_{\rm NN}} = 200$ GeV using the same settings in the AMPT model.

\label{AMPT}

\subsection{Elliptic flow}
\label{anisotropicflow}
In non-central heavy-ion collisions, finite azimuthal momentum space anisotropy is observed which could be expressed as the Fourier decomposition of azimuthal momentum distribution of particles in an event as,
\begin{eqnarray}
\frac{{\rm d}N}{{\rm d}\phi} = \frac{1}{2\pi}\Bigg[1+\sum_{n=1}^{\infty}2v_n \cos{\left(n(\phi-\psi_n)\right)}\Bigg].
\label{eq1}
\end{eqnarray}
Here, $v_n = \langle \cos{[n(\phi-\psi_n)]}\rangle$ is the $n^{\rm th}$ order harmonic flow coefficient, $\phi$ is the azimuthal angle and $\psi_n$ is the $n^{\rm th}$ harmonic symmetry plane angle~\cite{Voloshin:1994mz}. Anisotropic flow may be a combined outcome of initial spatial anisotropy of the nuclear overlap region, the transport properties and the equation of states of the produced system~\cite{Ollitrault:1992bk,ALICE:2016ccg}. Due to the almond-shaped ellipsoidal nuclear overlap region, the dominant contribution to anisotropic flow comes from the elliptic flow \textit{i.e.} the second order coefficient ($v_2$) in Eq.~\eqref{eq1}. To calculate the elliptic flow coefficients event-by-event, we have followed the event plane method~\cite{Masera:2009zz}. In AMPT simulation, there is a provision of making the reaction plane angle $\psi_n = 0$, although it is non-trivial in experiments. From this, one can obtain the elliptic flow coefficients as $v_n = \langle \cos{[n(\phi)]}\rangle$. The average is taken over all the chosen charged particle tracks for an event. For this study, we have used a fixed reaction plane angle, \textit{i.e.} $\psi_n = 0$ for both training and testing data sets. 


\section{Machine Learning based regression}
\label{ML}
In this section, we discuss the detailed analysis procedure including the input and output observables, the DNN architecture, training, evaluation, and quality assurance. The estimation of systematic uncertainties is also covered.

Computers require a specific algorithm to perform a task in which the solution to the problem is written in a top-to-down approach and the control flows accordingly resulting an outcome. Yet, most of the problems come with no standard predefined set of rules to develop the algorithm that can solve them. Another direction is the high-complexity non-linear problems, where linear-based numerical methods usually fail. In such cases, Machine Learning with smart algorithms such as the Boosted Decision Trees (BDT), Deep Neural Network (DNN), Generative Adversarial Network (GAN) \textit{etc.} could help the machine learn from the data through a process called training. ML is the branch of Artificial Intelligence (AI) that gives the ability to the computers to learn correlations from data components. In the field of astronomy, DNN models have been used to map complex non-linear functions by using simulated data~\cite{Ribli:2018kwb}.
This ability could be exploited to train ML-models to look for the hidden physics laws that govern particle production, anisotropic flow, spectra \textit{etc.} in heavy-ion collisions. A small attempt is made here to estimate the elliptic flow, $v_2$, by using powerful statistical tools implemented in DL algorithms to reproduce the observable of interest. For this purpose, we suggest the implementation of a feed-forward DNN in ML framework similarly as in Refs.~\cite{Monk:2018,Zhao:2021yjo,Biro:2021zgm}. 
Generally, a heavy-ion collision event produces multitude of particles in the final state. Each particle interacts with designated detectors to leave a track. These detectors can gather track information such as the phase space observables i.e. $p_{\rm T}, \eta, \phi$, charge \textit{etc.} and help in identifying the particles. To estimate the elliptic flow coefficient ($v_2$) on an event-by-event basis, we propose to train the machine with various particle kinematic properties. As the number of tracks varies from event to event, it makes the case a bit tricky to follow the conventional matrix based input space of a fixed order in a feed-forward DNN model to map the single output observable, here the $v_2$ and the energy or multiplicity scaling of the $v_2$. We propose an alternative yet convincing way to deal with this problem which is described below. 

\subsection{Input to the Machine}
\label{IO}

In this study, to train a DNN based ML-regression model for the target (output) variable $v_2$, binned $(\eta {\rm-}\phi)$ coordinate space for all charged hadrons in an event, has been taken as the primary input space. Here, $\eta \in [-0.8,0.8]$ and $\phi \in [0,2\pi]$. The bin number is chosen through a proper evaluation, which is discussed in the upcoming Section~\ref{QA}. To add further kinematic information necessary for estimation of $v_2$, we have included three secondary layers to the $(\eta {\rm -}\phi)$ space. The three different layers carry different weights as additional input for the ML-model. These three layers are weighted by the transverse momentum ($p_{\rm T}$), mass of the charged particle and $\log(\sqrt{s_{\rm NN}/s_0})$, a term related to the center-of-mass energy, where $\sqrt{s_0} = 1~ \rm GeV$ makes $\sqrt{s_{\rm NN}/s_0}$ unit-less. Figure~\ref{fig1} shows the three layers of the weighted $(\eta {\rm -}\phi)$ space having $32\times32$ bins each for a single minimum bias Pb-Pb collision at $\sqrt{s_{\rm NN}} = 5.02$~TeV from AMPT model simulation. Different colours are used to represent the different layers. The $p_{\rm T}$, mass and energy weighted plots are shown in blue, red and green colours respectively. The number of features, thus, would be $32\times32\times3=3072$. These feature values for an event could be extracted as the bin content after filling the respective two dimensional histograms with the tracks as shown in Fig.~\ref{fig1}. Thus, all the kinematic information from an event is mapped to these 3072 features which serve as the input neurons to  DNN. Now, the input space with a fixed feature set (matrix of dimension $1\times3072$) is ready to be used for training. We have considered all charged particles with transverse momentum cut $0.2 < p_{\rm T}<5.0$ GeV/c in pseudorapidity, $|\eta|<0.8$ for the training of the minimum bias ML-regression model. Before moving to the next step, which is the model design and training, the input array is normalised using the {\it L2-Norm} which makes the square root of sum of squares of the elements in the array equals to one. This way, the raw feature vectors are transformed into a more machine friendly representation by standardization of the data. It helps in faster convergence of the regression estimators.   This step also ensures that the coefficients of the model are small and in return, the model is less complex.

\begin{figure}[ht!]
    \includegraphics[scale=0.45]{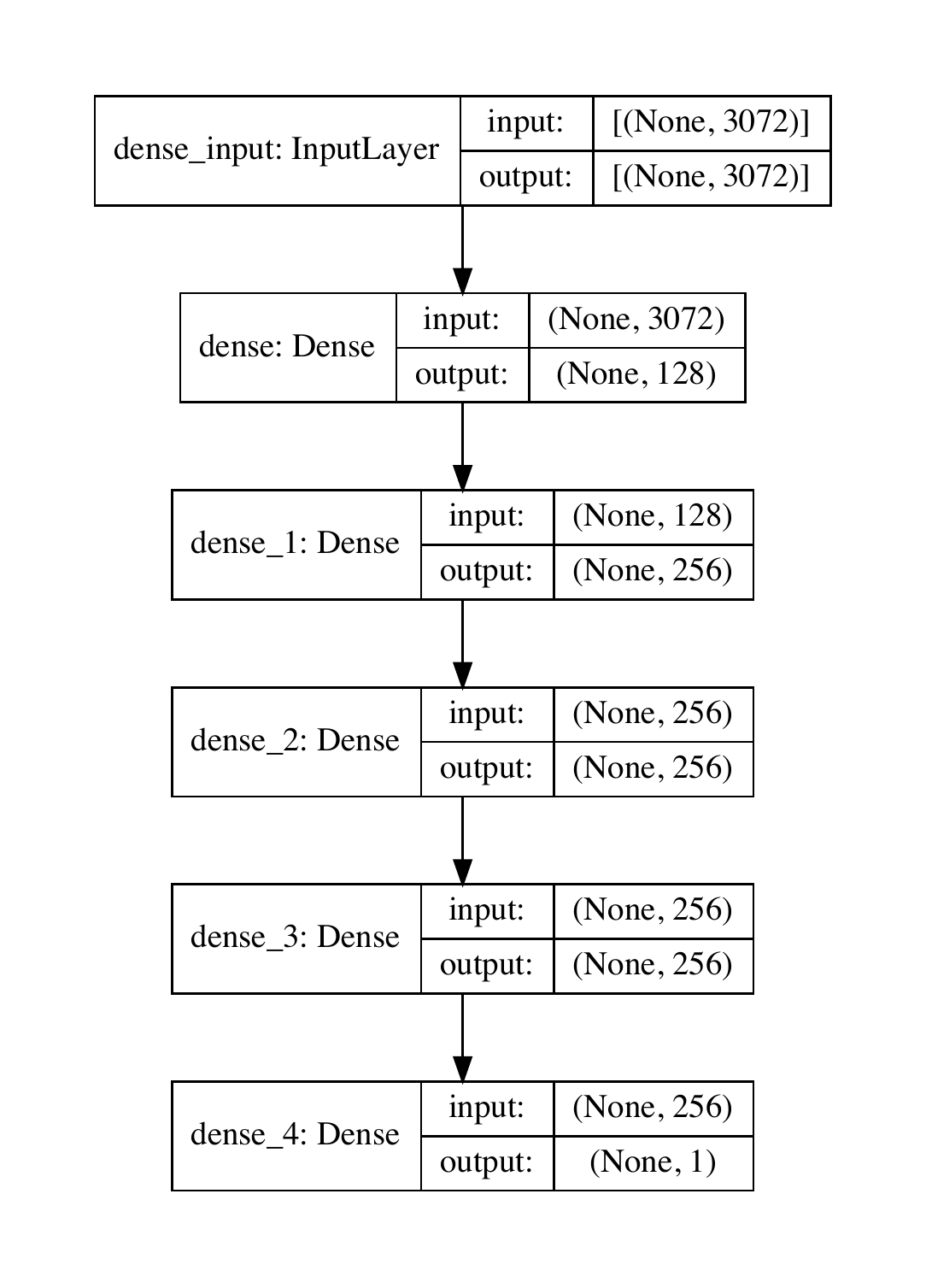}
    \caption{Schematic diagram depicting the DNN architecture used in this work.}
    \label{fig2}
\end{figure}

\subsection{Deep Neural Network}
\label{DNN}

Once being said about the input and output observables, now it is time to discuss the ML-regression algorithm. In high energy nuclear and particle physics, two of the most preferred ML algorithms are BDT and DNN.  The DNN is a powerful tool in Machine Learning and has been applied to numerous problems in HEP such as classical papers~\cite{Denby:1987rk,Lonnblad:1990bi}, jet tagging~\cite{deOliveira:2015xxd,Komiske:2016rsd,Chen:2019uar}, PID and track reconstruction~\cite{Farrell:2018cjr, Belayneh:2019vyx, Goncharov:2021wvd} and heavy-ion physics~\cite{Pang:2016vdc,Zhao:2021yjo,Biro:2021zgm,Xiang:2021ssj}. The interested readers may refer~\cite{Feickert:2021ajf} which contains an excellent collection of ML papers in particle physics, cosmology and beyond. A neural network is inspired from the biological neurons in animal brains where information is processed and communicated as signals through a proper pathway of neurons to realize an action or take a decision. DNN has several components. Starting from the input layer, where all the features are present which needs to be mapped to the output layer. The exact mapping function is not known and DNN is trained to learn the mapping by itself. The network consists of further intermediate (hidden) layers with different sets of nodes and finally the output layer. Each of the layers consisting of several nodes are connected to a subset of nodes in the next layer. If all the nodes of the previous layer connects each of the nodes of the present layer, then it is called a dense layer. A network consisting of several hidden dense layers is called as a deep network. The connection between two nodes in any adjacent layer is made mathematically with some weights and biases.

The existing problem of estimating $v_2$ from particle kinematic properties is of supervised regression kind. To learn the mapping function from the data, the neural network is shown with many events consisting of input (3072 features) and the corresponding true value of the output ($v_2$) for that event. A loss function evaluates the difference between the true value from simulation and the network output. The minimization of the loss function is performed through an optimizer algorithm which basically updates the weights and biases at each node at every stage of the training. An activation function is used to introduce nonlinearity into the model. This step is one of the crucial aspects of the network. Together with the linear transformations carried by the well optimised weights and biases, and the nonlinear effect of the activation function, a DNN can approximate solutions to complex nonlinear mapping functions. 

The Deep Neural Network used in this regression problem is shown pictorially in Fig.~\ref{fig2}. The network begins with the input feature layer which is mapped to the first dense layer with 128 nodes. It has further three hidden dense layers namely {\it dense\_1, dense\_2 and dense\_3} each having 256 nodes. For the input and hidden layers, rectified linear unit (ReLU) activation function is used~\cite{ReLU} which is of the form ${\rm ReLU(x) = max\{0,x\}}$. The output dense layer named as {\it dense\_4} consists of a single node as the $v_2$ and the linear activation function is used for this layer. The network is trained with the {\it adam} algorithm~\cite{Kingma:2014vow} as the optimizer with mean-squared-error {\it (MSE)} loss function. All the layers are dense layers.

\begin{figure}[ht!]
    \includegraphics[scale=0.35]{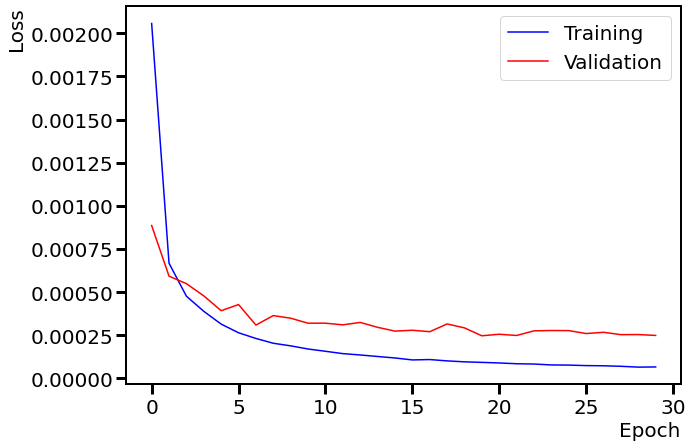}
    \caption{(Color online) Evaluation of mean squared loss using the \textit{adam} optimiser as a function of epoch size in the DNN during the training and validation for Pb-Pb collisions at $\sqrt{s_{\rm NN}} = 5.02$~TeV (min. bias) collisions from AMPT model.}
    \label{fig3}
\end{figure}

Deep Neural Network also faces the standard over-fitting issue like every other ML algorithm. Over-fitting is the scenario where the model picks up super-fine details of the training data set but it performs poorly with the validation set. A properly trained model should be stable over a large set of the training data and achieve minimum difference between the training and validation loss without compromising the accuracy of the prediction. To tackle the over-fitting issue, we have tried (i) dropout technique~\cite{Srivastava:2014kpo} and (ii) {\it L2- regularization}~\cite{l2regularization}. These methods, although helped in mitigating the over-fitting issue, yet drove the network far from accurate prediction, hence not used. Finally, a simple early stopping mechanism is used to solve this problem. The training is stopped if over-fitting is observed over a specified patience level. In this case, we have used a maximum of 60 epochs with 32 batch size and an early stopping patience level of maximum 10 epochs. The training is done with around 200K minimum bias events of  Pb-Pb collisions at $\sqrt{s_{\rm NN}} = 5.02$~TeV and 20\% events are used for validation. In the text, hereafter, the proposed DNN model refers to this model trained with minimum bias collision data only.

\begin{figure}[ht!]
    \includegraphics[scale=0.45]{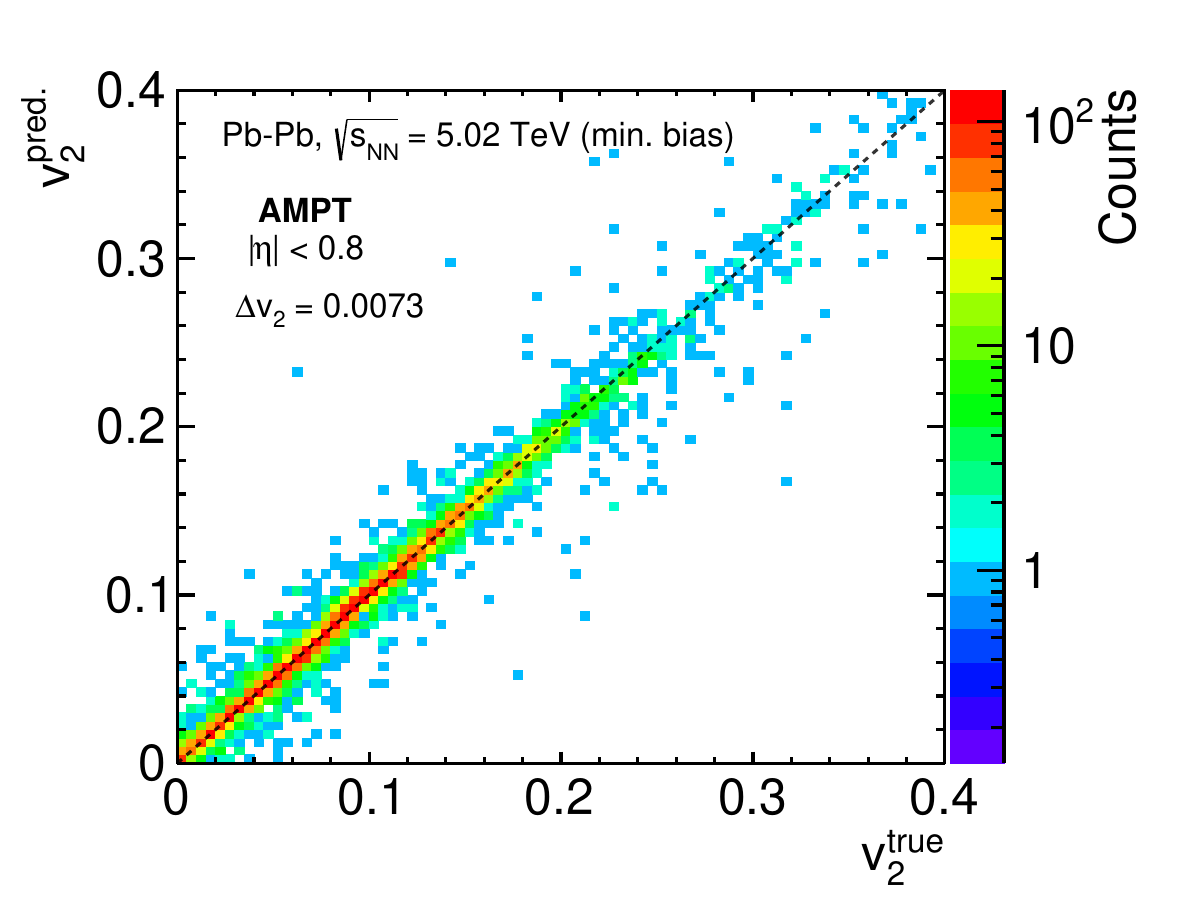}
    \caption{(Color online) DNN predictions versus the true values of elliptic flow coefficient ($v_2$) for 10K minimum bias events of Pb-Pb collisions at $\sqrt{s_{\rm NN}} = 5.02$~TeV from AMPT model.}
    \label{fig4}
\end{figure}

Figure~\ref{fig3}, shows the training and validation performance of the DNN model by evaluating the mean-squared-error {\it (MSE)} loss as the function of epoch size. The training and validation curves show sharp decrease with an increase in epoch. The interesting thing here is to note that, after a certain epoch size, the training is stopped with the early stopping callback to ensure there is minimal over-fitting. The difference between the validation and training loss is of the order of $10^{-4}$ which can be taken as a reasonably good training. The DNN model is now frozen and used for predictions. Figure~\ref{fig4} shows the predictions of $v_2$ from the model versus the true values from the simulation for 10K minimum bias Pb-Pb collisions at $\sqrt{s_{\rm NN}} = 5.02$~TeV from AMPT model. The mean-absolute-error (MAE) for $v_2$ which is defined in Eq.~\ref{eq2} is found to be $\Delta v_2 = 0.0073$. The predicted $v_2$ has a very good agreement with the true $v_2$ as the $v_2^{\rm true} = v_2^{\rm pred.}$ straight line shown in dashed black line is seen to be nicely populated in Fig.~\ref{fig4}.

\begin{eqnarray}
\Delta v_2 = \frac{1}{N_{\rm events}}\sum\limits_{n=1}^{N_{\rm events}}|v_{2_{n}}^{\rm true}-v_{2_{n}}^{\rm pred.}|
\label{eq2}
\end{eqnarray}

\subsection{Quality assurance}
\label{QA}  
    The selection of a bin number (grid dimension in Fig.~\ref{fig1}) for the two dimensional $(\eta {\rm -}\phi)$ space is done by training the model with input of varying bin numbers. The performance of the DNN with different number of bins in the input space has been listed in Table~\ref{tab1}. With $8\times8$, $16\times16$, $32\times32$ and $64\times64$ bins, the model is evaluated with 50K training events and 5K testing events with the exact settings for the DNN mentioned in Section~\ref{DNN}. From the performance of the model, it is seen that input space with both $8\times8$ and $16\times16$ number of bins, perform poorly with testing mean-absolute-error (MAE) being 0.0292 and 0.0171 respectively. For $32\times32$ bins, the model performed decently with testing MAE = 0.0102. Training with even higher bins in the input space \textit{i.e.} $64\times64$ not only renders the training slower as seen from the very high $\frac{\rm CPU~time}{\rm epoch} \approx 6~{\rm sec}$ but also gives even worse testing MAE = 0.0113. In this case, the number of trainable parameters of the DNN becomes too high ($\approx 1.7~{\rm M}$) which naturally slows down the training process. Evidently, $32\times32$ number of bins is taken as an ideal input size with respect to prediction accuracy and efficient training time. The specifications of the CPU used for the performance are as follows. The CPU type is Intel(R) Core(TM) i5-8279U (Released Q2'19) which has four cores (eight threads) clocked at base frequency 2.40 GHz and has a max turbo boost frequency of 4.10 GHz~\cite{cpu}. The system has 8 GB of LPDDR3 RAM clocked at 2133 MHz. The training is done with batch size fixed at 32. It could be noted here, due to the stochastic nature of the training process and type of CPU used, slight discrepancy in the performance of the DNN is expected. 
 
 \begin{figure}[ht!]
    \includegraphics[scale=0.45]{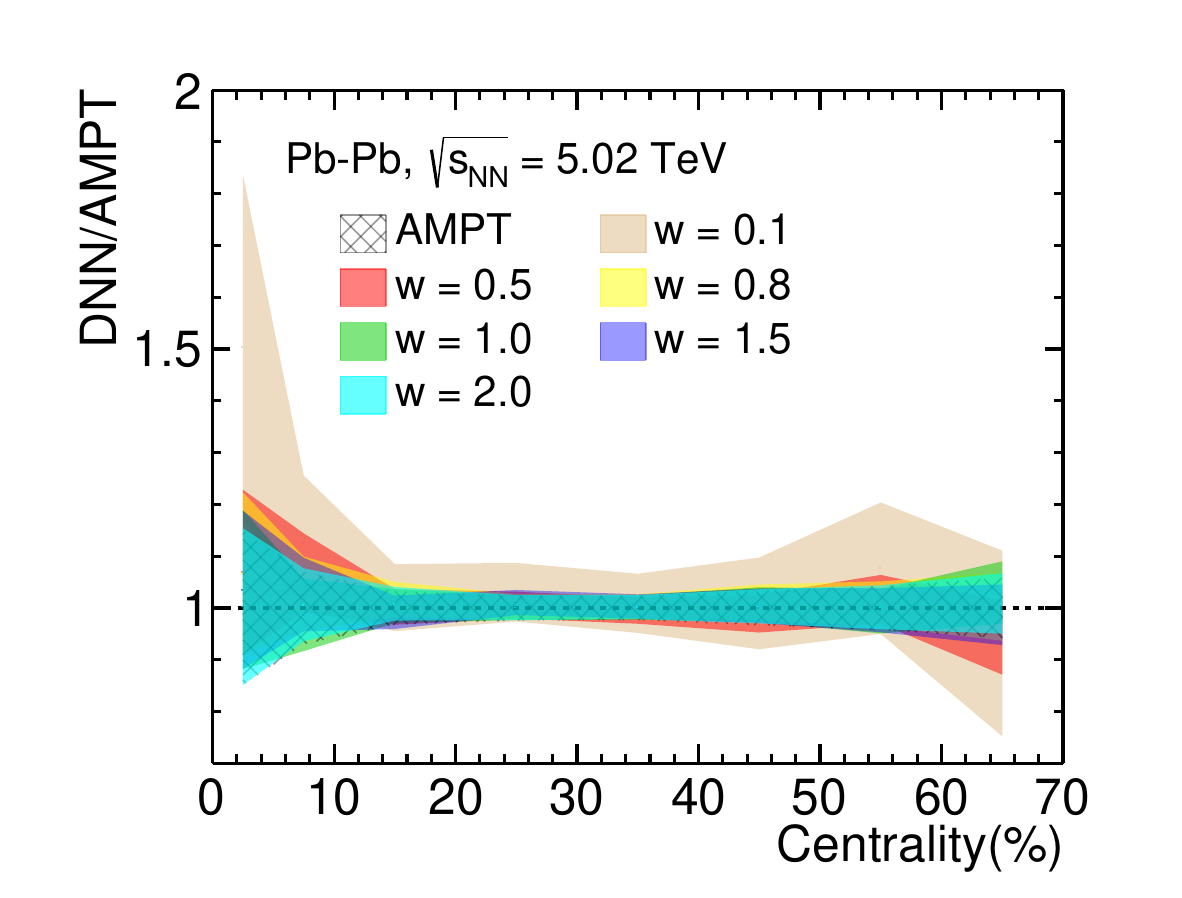}
    \caption{(Color online) Noise sensitivity test of the DNN model with respect to simulated values from AMPT. Smaller value of $w$ defines greater amount of noise in the dataset.}
    \label{fig5}
    \end{figure}

\begin{table}[ht!]
\begin{tabular}{|c|c|c|c|c|c|}
\hline
\begin{tabular}[c]{@{}c@{}}Bin \\ size\end{tabular} & \begin{tabular}[c]{@{}c@{}}Input \\ neurons\end{tabular} & MAE    & Epoch & $\frac{\rm Time~(sec)}{\rm Epoch}$ & \begin{tabular}[c]{@{}c@{}}Trainable \\ parameters\end{tabular} \\ \hline
$8\times8$                                          & 192                                                      & 0.0292 & 18    & 1.679                      & 189,569                                                         \\ 
$16\times16$                                        & 768                                                      & 0.0171 & 28    & 1.909                      & 263,297                                                         \\ 
$32\times32$                                        & 3072                                                     & 0.0102 & 30    & 2.684                      & 558,209                                                         \\ 
$64\times64$                                        & 12288                                                    & 0.0113 & 60    & 6.001                      & 1,737,857                                                       \\ \hline
\end{tabular}
\caption{Performance of the DNN model with different bin numbers in the $(\eta {\rm -} \phi)$ space for training with 50K events and testing with 5K events.}
\label{tab1}
\end{table}

\begin{figure*}[ht!]
    \includegraphics[scale=0.29]{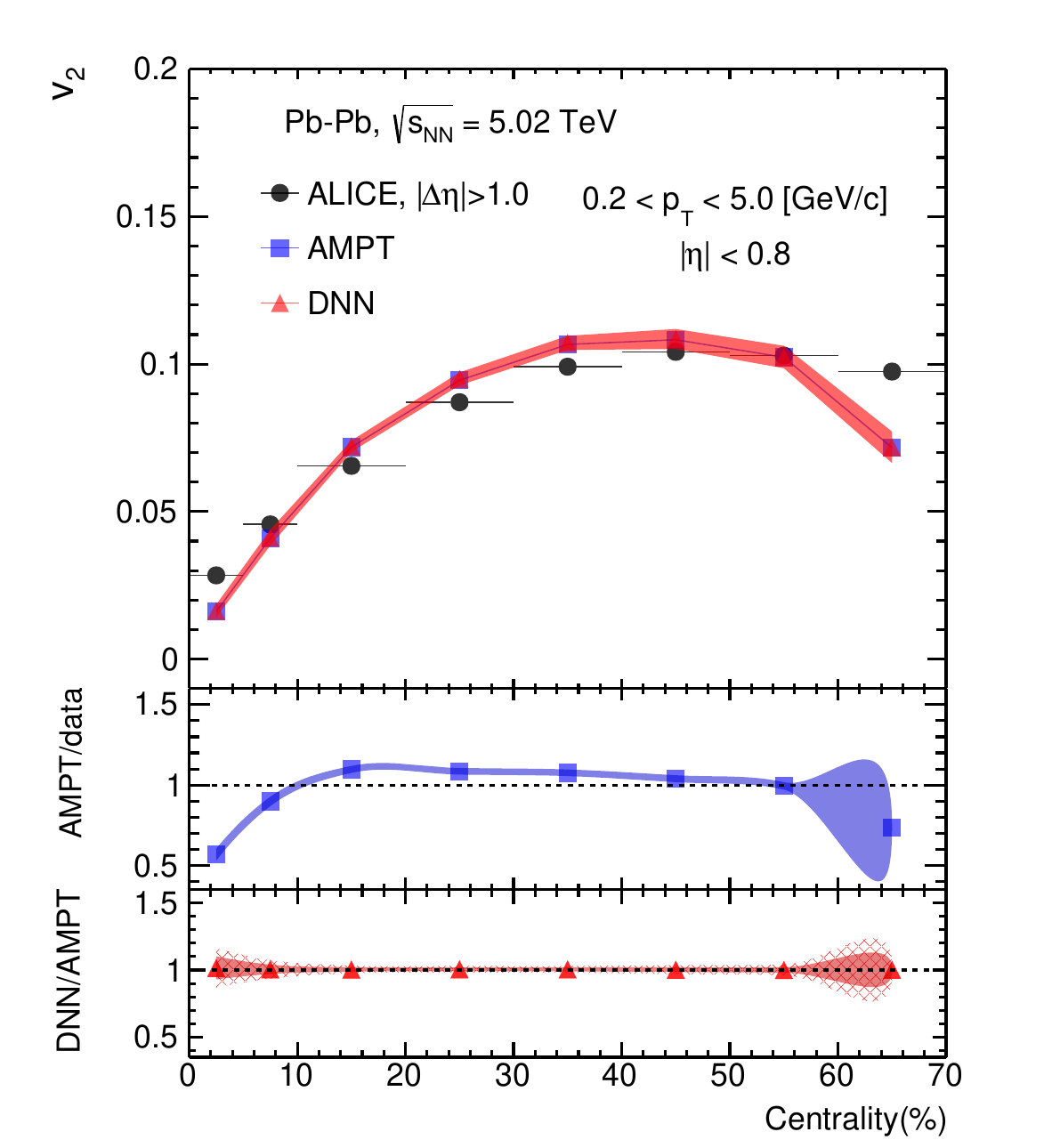}
    \includegraphics[scale=0.29]{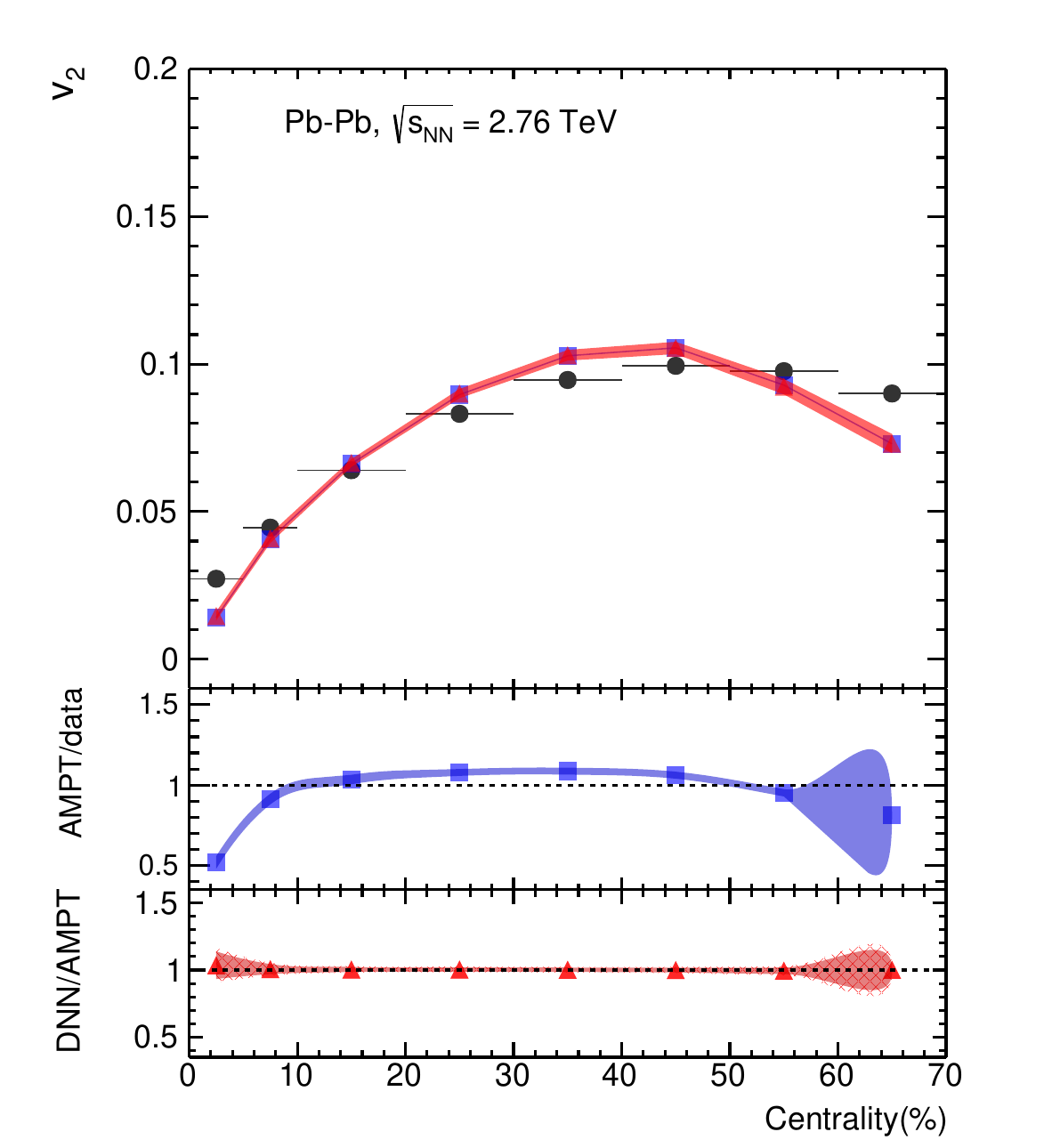}
    \includegraphics[scale=0.29]{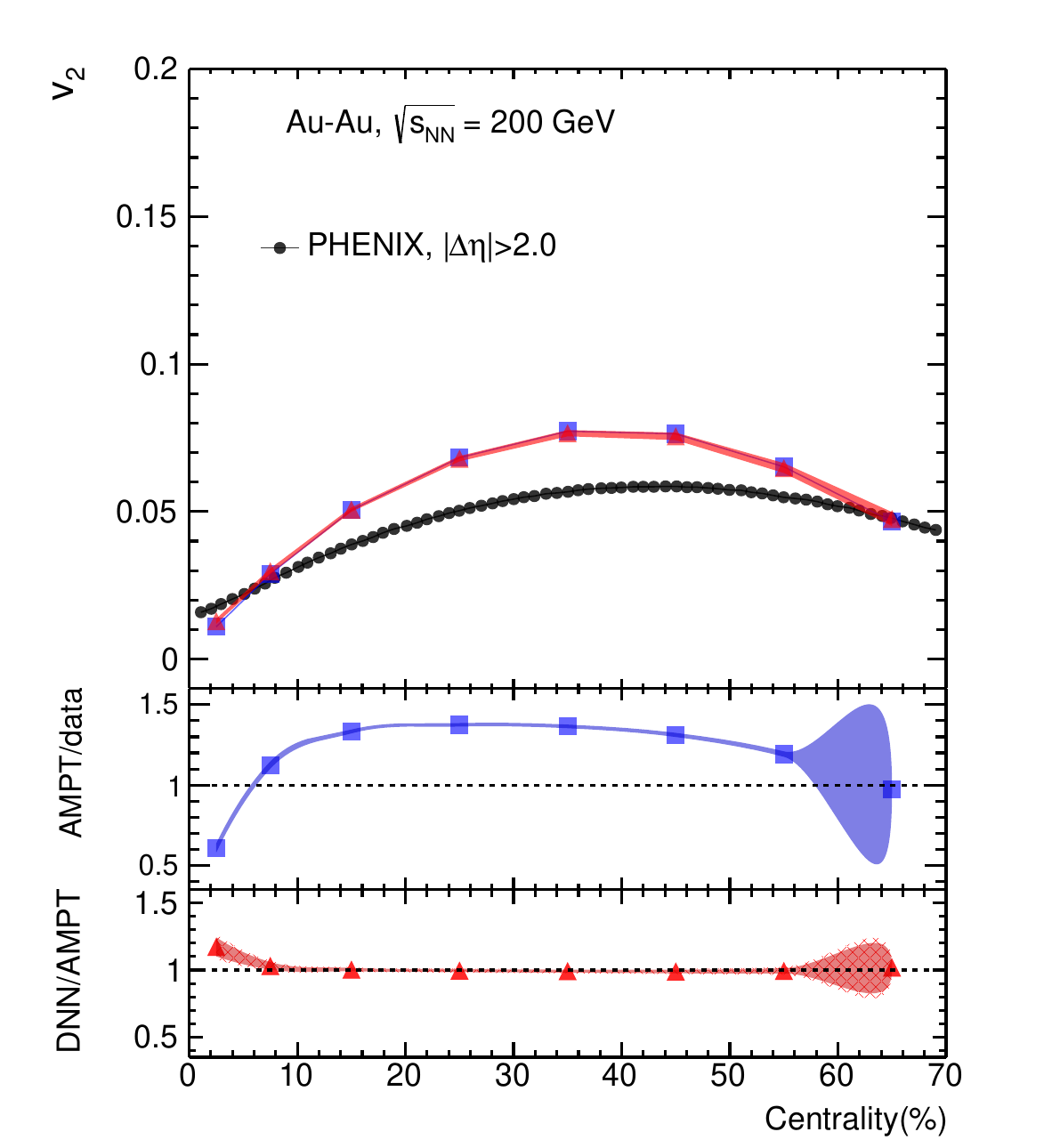}
    \caption{(Color online) Estimation of $v_2$ for different centrality classes in Pb-Pb collisions at $\sqrt{s_{\rm NN}} = 5.02$~TeV, $2.76$~TeV and Au-Au collisions at $\sqrt{s_{\rm NN}} = 200$~GeV from AMPT model using DNN. Data from ALICE and PHENIX for comparison.}
    \label{fig6}
\end{figure*}

Deep Neural Networks are known to be quite stable and robust to random noise or fluctuations in the data~\cite{Komiske:2016rsd}. To check the noise sensitivity of prediction of our working DNN model, we proceed with this simple test. In this study, there are 3072 input features ($F_{i,j}$) for each event, $i$ is the $i^{\rm th}$-event and $j$ is the $j^{\rm th}$-feature. To introduce noise to each of these features, one must need to evaluate the standard deviation ($\sigma_j$) associated with each of these features ($j$). Following the central limit theorem, each of these features must describe a Gaussian density function. We assume that the noise ($X_{i,j}$) introduced to each feature value ($F_{i,j}$) should be proportional to a random number between $(-\sigma_j, \sigma_j)$. Hence, we randomly select a number $X_{i,j}$, such that $X_{i,j} \in (-\sigma_j, \sigma_j)$. Now, we introduce $w$ which is the weight factor. The weight factor helps to define the magnitude of the noise. For each feature value $F_{i,j}$, we then add the noise $X_{i,j}/w$, such that $F_{i,j} = F_{i,j} + X_{i,j}/w$. A higher value of $w$ would correspond to lower noise and {\it vice-versa}. Again, a smaller value of $w$ would broaden the width of the feature distribution and hence changing its true shape and evidently affecting the DNN input. This can be treated as an imperfect simulation dataset. In Fig.~\ref{fig5}, we have evaluated the noise sensitivity of the DNN model by taking different weights. The ratio of DNN to AMPT shows the degree of agreement between the simulation and the machine prediction. For perfect prediction, the ratio should be exactly equal to one. The width of each band shows the statistical uncertainty associated with it for each centrality class. As we can clearly observe, the smaller value of $w$ gives larger noise to the dataset, resulting in a greater deviation from the true value. Whereas a reasonable amount of noise in the dataset does not affect the prediction of the DNN much. This test helps us understand the noise sensitivity of the DNN model. We can conclude that the DNN model is not that sensitive to random noise and hence pretty stable and accurate. The deviation between the model prediction for a fair simulation and a noisy simulation with a certain $w$ could constitute the systematic uncertainty of the method. We consider $w = 0.5$ which is reasonable for estimation of the systematic uncertainties which are included in the centrality wise predictions.

Note, similar supervised regression problems on estimation of impact parameter in high-energy heavy-ion collisions have been investigated with convolutional neural network (CNN) based models~\cite{Xiang:2021ssj,Zhang:2021zxd}, and PointNet model~\cite{OmanaKuttan:2020brq}, where authors use image-like inputs for RHIC and lower energies. For our case, we have used a matrix based input derived from the image-like histograms, and it is found that the proposed DNN model is quite efficient in this study. We also recommend to try CNN and PointNet model based approach which are specialized algorithms for handling image-like inputs directly for mapping the flow coefficients. It can be taken as an outlook for the present work, and it is not covered here.

\section{Results and Discussions}
\label{results}
We have made an attempt to implement the DNN in the ML framework in heavy-ion collisions using a novel technique to include the particle kinematic properties to estimate the elliptic flow coefficient, $v_2$. The training is done with minimum bias Pb-Pb collisions at $\sqrt{s_{\rm NN}} = 5.02$ TeV generated from AMPT and the model is applied to successfully predict centrality-wise $v_2$ for Pb-Pb collisions at $\sqrt{s_{\rm NN}} = 5.02,~2.76$ TeV and Au-Au collisions at  $\sqrt{s_{\rm NN}} = 200$ GeV, shown in Fig.~\ref{fig6}. We have compared DNN predicted $v_2$ values with the experimental observations at the LHC (left and middle)~\cite{ALICE:2016ccg} and the RHIC (right)~\cite{PHENIX:2018lfu}.

One can clearly see from the AMPT to data ratio plots for the LHC energies that, with the current settings, AMPT reproduces the data very nicely. However, there are some level of discrepancies for the most central(peripheral) cases. It should also be noted here that the estimation of $v_2$ by ALICE using $|\Delta \eta| > 1$ could have some level of non-flow effects. In addition, different method of flow estimations also introduce a degree
of uncertainty/mismatch. For RHIC energy, the current settings over-estimate $v_2$ as compared to the PHENIX result. This could be avoided using a specific tune valid for the top RHIC energies as proposed by the authors of AMPT~\cite{Lin:2014tya}. Here, we stick to the LHC tunes of AMPT mentioned in Section~\ref{AMPTModel}, as the primary focus of the paper is towards the study of DNN implementation in heavy-ion collision system. 

We find a very good agreement between the $v_2$ values predicted from the proposed DNN based model and the true $v_2$ values calculated from AMPT for all centrality classes for Pb-Pb collisions at $\sqrt{s_{\rm NN}} = 5.02,~2.76$ TeV and Au-Au collisions at  $\sqrt{s_{\rm NN}} = 200$ GeV. The agreement can be seen in the DNN to AMPT ratio plot. The estimation of systematic uncertainty is already discussed in the earlier section. As one can see in the top panel of Fig. \ref{fig6}, {\it i.e.} $v_2$ {\it vs.} centrality plots, the quadratic sum of statistical and systematic uncertainty are shown in the red solid band. However, in the lower bottom panel {\it i.e.} DNN to AMPT ratio plots, the statistical uncertainty is shown in solid red band and the systematic uncertainty is shown in dashed red band. 

\begin{figure}[ht!]
    \includegraphics[scale=0.45]{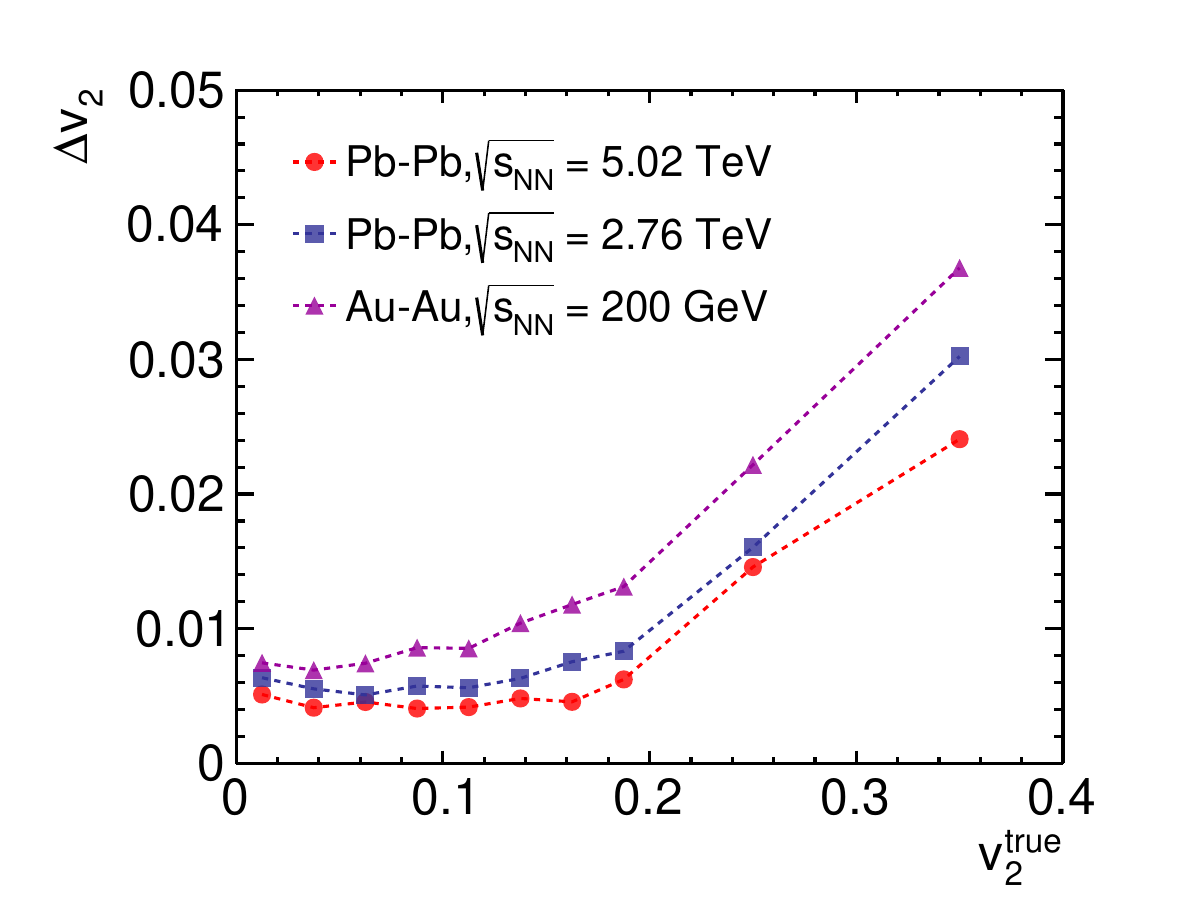}
    \caption{(Color online) Performance of the DNN model trained on Pb-Pb collisions, $\sqrt{s_{\rm NN}} = 5.02$ TeV (min. bias) applied to different collision energies for different regions of $v_2^{\rm true}$. $\Delta v_2$ is the mean-absolute-error for the given bin of $v_2^{\rm true}$.  }
    \label{fig7}
\end{figure}

By training the DNN model with minimum bias Pb-Pb collisions at $\sqrt{s_{\rm NN}} = 5.02$ TeV, we allow the machine to learn physics for a larger and complex system. The striking feature of the model is that, with the minimum bias training, it is perfectly capable of reproducing the centrality-wise $v_2$ values not only for the trained energy but also for lower energies {\it i.e.} Pb-Pb collisions, $\sqrt{s_{\rm NN}} = 2.76$ TeV and Au-Au collisions, $\sqrt{s_{\rm NN}} = 200$ GeV. This is well understood from the bottom ratio plots. One can interpret these results as the fact that the proposed DNN model with the selected input parameters learns and preserves the centrality and energy dependence of elliptic flow.

Variation of mean-absolute-error, $\Delta v_2$, is shown for different regions of $v_2^{\rm true}$ in Fig.~\ref{fig7}. This plot is a visual representation of performance of the proposed DNN model trained on minimum bias Pb-Pb collisions at $\sqrt{s_{\rm NN}} = 5.02$~TeV, that predicts the $v_2$ values for various collision energies at RHIC and LHC for different regions of $v_2^{\rm true}$. For the region $v_2^{\rm true} <$ 0.2, variation in $\Delta v_2$ is negligible at LHC energies. However, a small rise in $\Delta v_2$ with increasing $v_2^{\rm true}$ is observed at RHIC energy. The maximum relative error ($\Delta v_2/v_2^{\rm true}$), in this case, is less than 6.0\%. For the region $v_2^{\rm true} > 0.2$, $\Delta v_2$ seems to increase sharply for both RHIC and LHC energies. This could be due to the fact that event statistics for higher elliptic flow values ($v_2^{\rm true} > 0.2$) are naturally less and the model sees less examples for this region in an inclusive and unbiased training data set. However, the maximum relative error in this region is found to be between (6-10)\% only, which reflects the overall satisfactory performance of the proposed DNN model in a wide range of $v_2^{\rm true}$.

\begin{figure}[ht!]
    \includegraphics[scale=0.35]{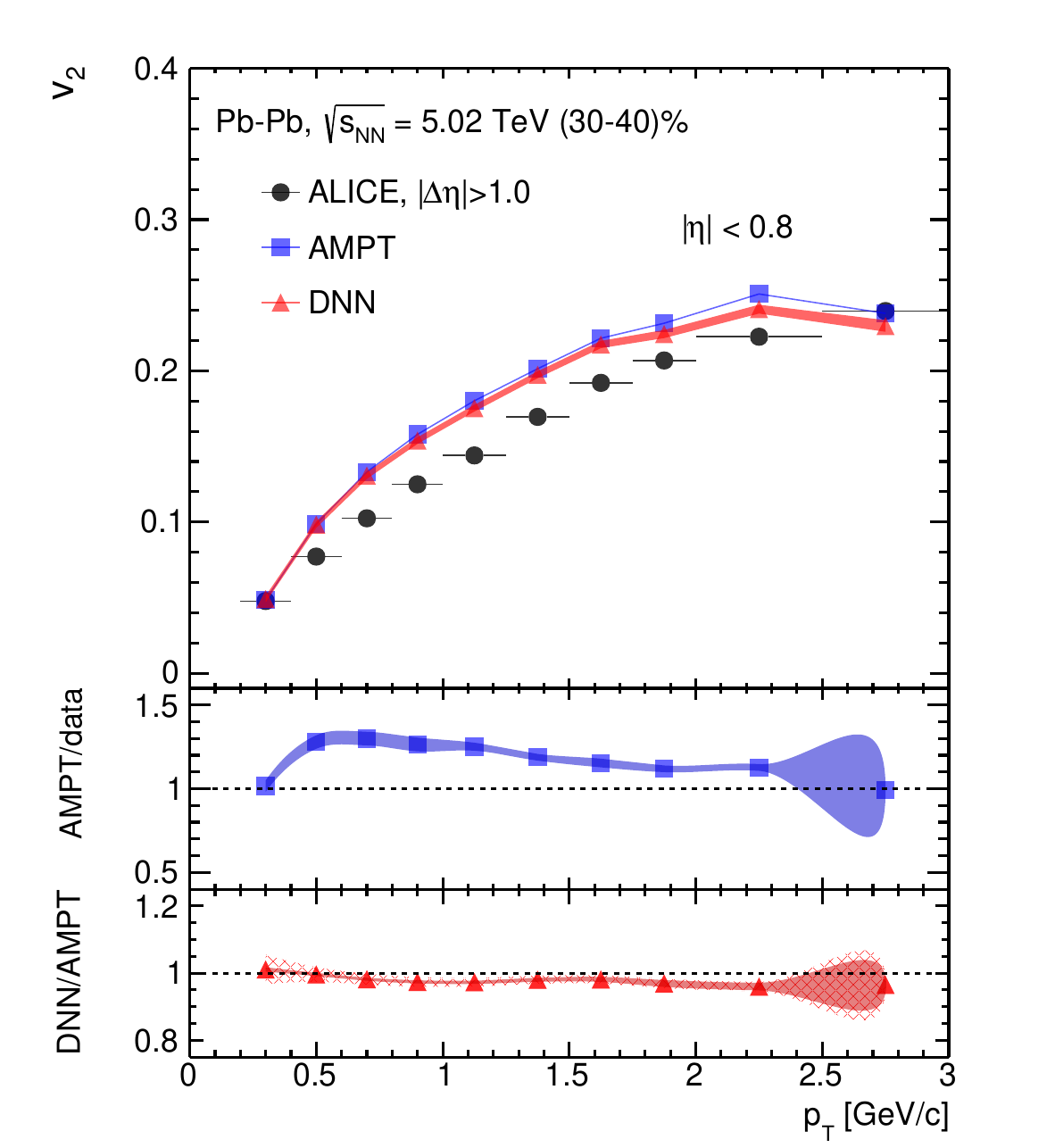}
    \caption{(Color online) Estimation of $v_2$ as a function of $p_{\rm T}$ for (30-40)\% centrality class in Pb-Pb collisions at $\sqrt{s_{\rm NN}} = 5.02$~TeV from AMPT using the DNN model. ALICE results for comparison.}
    \label{fig8}
\end{figure}

Finally, Fig.~\ref{fig8} shows the $v_2(p_{\rm T})$ for (30-40)\% central Pb-Pb collisions at $\sqrt{s_{\rm NN}} = 5.02$ TeV predicted from the DNN model. For comparison, we present ALICE measurement~\cite{ALICE:2016ccg} and $v_2^{\rm true}$ calculated from AMPT simulations for the same centrality class. The DNN prediction has a similar trend of $v_2(p_{\rm T})$ but it over-predicts the ALICE data in intermediate $p_T$, while it explains the data at low- and high-$p_T$. On the other hand, predicted $v_2$ has very good agreement with the $v_2^{\rm true}$ from AMPT for the full range of $p_{\rm T}$. This could be seen from the DNN to AMPT ratio plot on the bottom panel which stays almost close to unity. The quadratic sum of systematic and statistical uncertainty is shown in the solid red band on the upper panel $v_2(p_{\rm T})$ plot. On the bottom panel, in DNN to AMPT ratio plot, the red band shows the statistical uncertainty and the dashed band shows the systematic uncertainty for this centrality class.

\section{Summary}
\label{summary}
In summary, we report the implementation of DNN model in ML framework with a novel technique to include particle kinematic properties in heavy-ion collisions and estimate the elliptic flow. The proposed DNN model uses the kinematic properties such as $p_{\rm T}$, mass and  $\log(\sqrt{s_{\rm NN}/s_0})$ -- a term related to center-of-mass energy as model input. These information are encoded as the secondary layers in ($\eta{\rm -}\phi$) primary input space. The model is trained with minimum bias Pb-Pb collisions at $\sqrt{s_{\rm NN}} = 5.02$~TeV from AMPT simulations. The  performance of the model is tested under random fluctuations in data and the systematic uncertainties are calculated. Further, the trained model is applied to predict centrality-wise evolution of $v_2$ for Pb-Pb collisions at $\sqrt{s_{\rm NN}} = 5.02,~2.76$ TeV and Au-Au collisions at  $\sqrt{s_{\rm NN}} = 200$ GeV. From the results, a very good agreement is observed between the model prediction and simulated true value. The proposed DNN model seems to learn and preserve the centrality and energy dependence of elliptic flow pretty nicely. On top of that, the same model is  successfully applied to study the $p_{\rm T}$ dependence of $v_2$. From the results, the proposed DNN model seems to preserve the $p_{\rm T}$ dependence of $v_2$ in heavy-ion collisions as well.

 It should be noted here that, the machine learning-based model, specially used for high-energy physics, becomes more useful when the Monte-Carlo simulations explain experimental data as closely as possible. The ML-based model could be more realistic by taking into account the detector resolution and noise during training, which hopefully contain both the correlated and uncorrelated noises. Authors are planning to investigate these issues in a future study.

\section*{Software}
Although there are plenty of software available for implementing a DNN model, we have specifically used Keras v2.6.0 Deep Learning API~\cite{keras} with Tensorflow v2.6.0~\cite{Abadi:2016kic} backend in Python, to implement the DNN model used in this work. We also found the scikit-learn ML framework very helpful~\cite{sklearn}.

\section*{Acknowledgements}
SP acknowledges the doctoral fellowship from UGC, Government of India. RS acknowledges the financial supports under the CERN Scientific Associateship and the financial grants under DAE-BRNS Project No. 58/14/29/2019-BRNS. ANM and GGB gratefully acknowledge the Hungarian National Research, Development and Innovation Office (NKFIH) under the contract numbers OTKA K135515, K123815 and NKFIH 2019-2.1.11-TET-2019-00078, 2019-2.1.11-TET-2019-00050 and Wigner Scientific Computing Laboratory (WSCLAB, the former Wigner GPU Laboratory). The authors gratefully acknowledge the MoU between IIT Indore and WRCP, Hungary under which this work has been carried out as a part of the techno-scientific international cooperation. The authors are thankful to Dr. Sushanta Tripathy for his careful reading of the manuscript and insightful comments.

\end{document}